\documentclass{article}
\usepackage{spconf,amsmath,graphicx}
\usepackage{amssymb}
\usepackage{multirow}
\usepackage{booktabs}
\usepackage{subfigure}
\usepackage{tabularx}

\newtheorem{remark}{Remark}

\title{model copyright protection in buyer-seller environment}
\name{Yusheng Guo, Nan Zhong, Zhenxing Qian, Xinpeng Zhang\thanks{Thanks to XYZ agency for funding.}}
\address{School of Computer Science, Fudan University, Shanghai, China}
\begin{document}
%
\maketitle
\begin{abstract}

Training a deep neural network (DNN) requires a high computational cost. Buying models from sellers with a large number of computing resources has become prevailing. However, the buyer-seller environment is not always trusted. To protect the neural network models from leaking in an untrusted environment, we propose a novel copyright protection scheme for DNN using an input-sensitive neural network (ISNN). The main idea of ISNN is to make a DNN sensitive to the key and copyright information. Therefore, only the buyer with a correct key can utilize the ISNN. During the training phase, we add a specific perturbation to the clean images and mark them as legal inputs, while the other inputs are treated as illegal input. We design a loss function to make the outputs of legal inputs close to the true ones, while the illegal inputs are far away from true results. Experimental results demonstrate that the proposed scheme is effective, valid, and secure.

\end{abstract}
\begin{keywords}
deep  neural  networks, model copyright protecting, cryptology, input-sensitive neural network.
\end{keywords}
\section{Introduction}\label{1}

Deep neural networks (DNNs) show prominent superiority in a large variety of fields, including self-driving cars \cite{onishi2019end}, facial recognition authorization \cite{parkhi2015deep}, object detection \cite{szegedy2015going}, etc.
Training neural network models is expensive, which relies on extensive datasets and computing resources.
However, the resources of ordinary institutions or individuals are not always abundant.
They prefer to buy the well-trained DNNs from the enterprises, e.g. Google.
DNNs are becoming a valuable software commodity.

Unfortunately, the buyer-seller process is not always executed in a trusted environment.
The DNNs are at risk of being stolen.
Once the DNNs are stolen, their commercial value will be seriously damaged, which damages the interests of both buyers and sellers. 
Therefore, a countermeasure for the stealing of DNNs is eagerly required.

Considering a buyer-seller scenario like the Microsoft Office and MATLAB, the buyers purchase the well-trained DNNs from the seller.
The seller publishes a DNN to a cloud server or website that anyone can access. And he sends secretly a key to the buyer for accessing DNN if he has paid for it.
Meanwhile, an adversary attempts to steal DNN for using free or reselling.

\begin{figure}[t]
  \centering
  \includegraphics[width=0.47\textwidth,trim=310 170 300 160,clip]{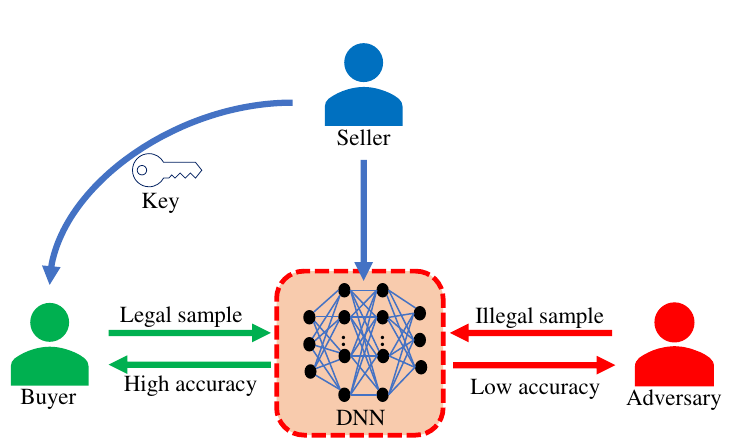}
  \caption{the framework of proposed scheme over images classification task.}\label{fig1.1}
\end{figure}

To prevent the DNNs from being stolen, the most straightforward approach is encrypting all parameters of DNN via conventional cryptography protocol such as RSA\cite{RSArivest1978method}, AES\cite{AESdaemen1999aes}, TDES\cite{TDEScoppersmith1996proposed}, and Twofish\cite{Twofishknudsen1998deal}.
The approaches are very secure, however, the computational complexity of decryption is generally not less than $\mathcal{O}(n)$.
Even in the best case, the computational overhead increases linearly with DNN size, which is not suitable for buyers with limited computing resources.
Some researchers apply the selective encryption to reduce the computational complexity of encryption and decryption \cite{chakraborty2020hardware,tian2021probabilistic}.
In \cite{chakraborty2020hardware}, the model can only be executed on trusted hardware.
The scheme in \cite{tian2021probabilistic} has to finding suitable nodes for encryption.
Meanwhile, the DNNs lose the inference ability once they are encrypted, which means that encryption is not perfect for model copyright protection in untrusted environments.
Model watermarking is also measures to protect neural network copyright \cite{guo2018watermarking,hitaj2018have,le2020adversarial,darvish2019deepsigns}.
Nevertheless, the buyer and seller cannot prevent the adversary from accessing the model when the DNNs are stolen.

In this paper, we propose a novel scheme based on the input-sensitive neural network (ISNN) to resist the model from leaking in the untrusted environment.
The main idea is to make a DNN very sensitive to input data.
Even if the two inputs are only slightly different, the outputs are not the same.
Fig.\ref{fig1.1} illustrates the framework of the proposed scheme over the images classification task.
The seller publicizes an ISNN in an untrusted environment and sends the key to the buyer over a secure channel.
The legal inputs involve stacking the clean sample with a specific perturbation that depends on the key, while the other inputs, including clean inputs, are treated as illegal inputs.
ISNN is credible if and only if the inputs are legal.

The accuracy of the legal inputs hardly decreases relative to the original task model.
However, the adversary cannot obtain satisfactory accuracy even he can access the entire parameters of ISNN.
In addition, the inference phase of ISNN can be executed in the untrusted environment without encryption and decryption.
Finally, we consider three potential attack methods, retraining attack, forging copyright attack, and reverse iterating attack to validate the security of the proposed scheme.
Experimental results demonstrate that the proposedscheme is effective, valid, and secure.

\section{Threat Model and Design Goals}\label{2}

\subsection{Threat Model}\label{2.1}

In an untrusted environment, the adversary can easily obtain the details of the DNN $\mathcal{F}$, including the model framework and the parameters of nodes $\Theta$.
The seller publishes the copyright information $I$.
We consider three attack modes, \emph{retraining attack}, \emph{forging copyright attack}, and \emph{reverse iterating attack}.

The retraining attack refers to that the adversary uses part of the dataset to retrain the DNN stolen.
For proving that the security of the proposed scheme is strong, we further improve the attack ability of the adversary by permitting the adversary to access the entire clean dataset $\mathcal{D}$ training.
However, the adversary does not know the algorithms to generate a legal dataset $D_T$.
He only retrains the ISNN using a clean dataset with the expectation of obtaining a model that can correctly classify clean samples.

In the forging copyright attack scenario, the adversary understands that the legal inputs are generated by adding a uniform noise to the clean samples.
He tried to steal the copyright of ISNN by injecting fake copyright information to displace $I$.
Specifically, the adversary forges a fake dataset $\widehat{D} $ to replace the legal dataset $D_T$ in the training process, where $\widehat{D}$ is generated by adding a uniform noise to $\mathcal{D}$.

The reverse iterating attack is based on the differentiability of DNN.
In the proposed scheme, the only difference between the buyer and the adversary is whether adds the correct disturbance to the input, seeing section \ref{3} for details.
Moosavi-Dezfooli et al.\cite{uap} proposed that adding universal perturbation to DNN can make a large number of samples misclassified, i.e. universal adversarial sample attacks.
If the adversary possesses a part of the training dataset, he may obtain the correct perturbation or perturbation with the same effect through the targeted universal adversarial sample attack. To reinforce the attack capability of the adversary, we permit the adversary to access the perturbation intensity $\epsilon$ and types (uniform, Gaussian, or impulse).

\subsection{Design Goals}\label{2.2}

We denote the accuracy $P$ of the protected DNN as the probability that the input sample is correctly classified. Mathematically,
\begin{equation}\label{eq2.2.1}
  P = \Pr[\mathcal{F}_{\Theta}(x) = y].
\end{equation}
where $x$, $y$, and $\mathcal{F}_{\Theta}$ is input image, the true label, and the DNN with parameters $\Theta$, respectively.
We also denote $P_{opt}$ is the optimal accuracy of the DNN model without protection measures. 
Baseline $\mathcal{T} = P_{opt}-10\%$ is used to evaluate the usability of the DNN. If the accuracy $P$ is below the baseline $\mathcal{T}$, it is considered useless.

To ensure delivery and usage of DNN in the untrusted environment, we propose the following design goals.
\emph{Verifiability:} The seller and buyer can verify the ownership of DNN even if the adversary modifies the parameters of DNN.\emph{Low-complexity:} The computational complexity for the buyer should satisfy $\mathcal{O}_{buyer} \ll  \mathcal{O} (n)$.
\emph{Flexibility:} Inference can be executed directly in the untrusted environment without decryption.
\emph{Crypticity:} Perturbation $\delta $ should be imperceptible. $\| \delta \| _{p} \leq \epsilon $ where $\| \cdot \| _{p}$ is $l_p$ norm.
\emph{Fidelity:} The accuracy of buyer $P_{buyer}$ over classification tasks should be kept above the baseline, i.e. $P_{buyer} < \mathcal{T}$.
\emph{Effectiveness:} The accuracy of the adversary $P_{adv}$ is significantly degraded. i.e. $P_{adv} > \mathcal{T}$.
\emph{Security:} The supremum of accuracy obtained by the adversary is lower than the baseline, in the threat model proposed, i.e. $\sup  P_{adv}^{att} \leq  \mathcal{T}$.

\section{proposed method}\label{3}

\begin{figure*}[htbp]
  \centering
  \includegraphics[width=1.00\textwidth,trim= 30 90 20 80,clip]{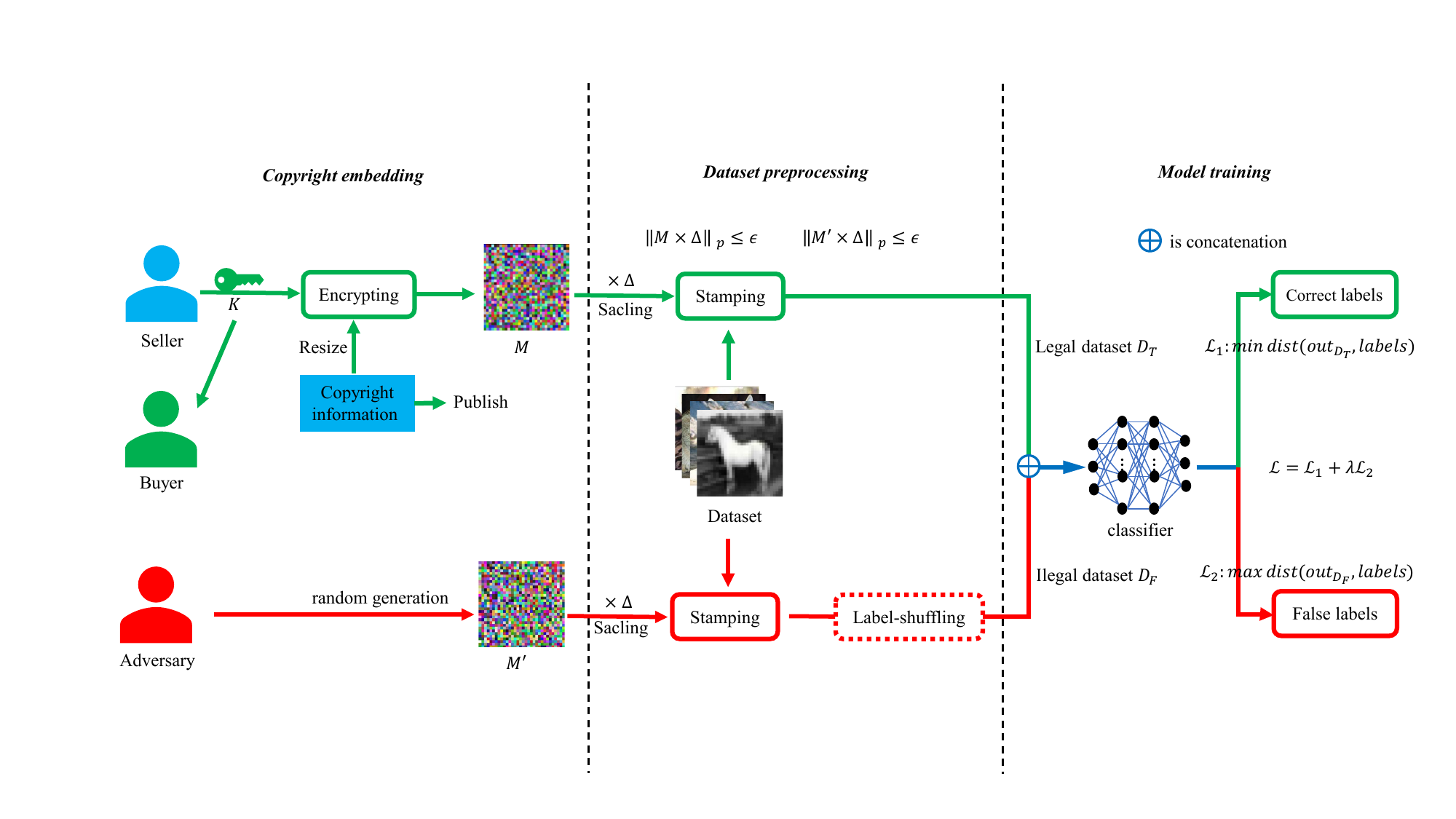}
  \caption{Structure of model protection scheme based on ISNN}\label{fig3.1}
\end{figure*}

Our core idea is that the model differentiates automatically between the buyer and adversaries based on input, instead of the encrypting model.
Overall, the operations of the proposed scheme can be roughly divided into upper and lower parts as illustrates in Fig.\ref{fig3.1}.
The main purpose of the upper part is to provide the highest possible accuracy for the buyer, and the bottom is aiming to slash the accuracy of the adversary.
Operations represented by dotted lines are optional.
The training process includes three modules according to the operation sequence \emph{Copyright embedding}, \emph{Dataset preprocessing}, and \emph{Model training}.
All operations are executed by the seller independently in the training phase, except the key generation.
Which requires the participation of the buyer to offers identity information.
The details of the four modules are described sequentially as follow:

\subsection{Copyright embedding}\label{3.1}

In the proposed scheme, the seller generates a 256-bit key $K$ and sends it to the buyer via relevant security technology, such as multi-party secure computing\cite{MPC}, homomorphic encryption\cite{HE}, and secret sharing\cite{SS}.
These techniques are mature and beyond the scope of this paper.
The security of copyright protection depends entirely on the key, hence $K$ must be strictly secret.

The seller records the copyright information $I$ as an image, then resizes it to match the input size of ISNN.
The copyright information $I$ is encrypted using the AES\cite{AESdaemen1999aes} encryption algorithm with the key $K$ then the ciphertext is denoted as $M$.
$M$ must be kept secret also.
Considering the security of AES, the adversary knows nothing about $K$ and $M$ from copyright information $I$,
There is virtually no difference between encrypting the copyright information with the false key and randomly generating a false $M'$.
In the training process of the proposed scheme, $M'$ is randomly generated rather than obtained by encrypting.

\subsection{Dataset preprocessing}\label{3.3}

Crypticity is necessary to prevent an adversary from obtaining the key by monitoring the input of the ISNN.
Therefore, we constrain the perturbation intensity according to the crypticity in \ref{2.2}.
The perturbation $\delta$ satisfies the following condition.

\begin{equation}\label{eq3.3.1}
  \delta = M \times \Delta,
  \parallel \delta\parallel_{p} \leq \epsilon,
\end{equation}
where $\epsilon$ is the perturbation intensity, $\Delta$ is the scale factor, and $\parallel \cdot\parallel_{p}$ is $l_{p}$ norm. In our scheme $p = \infty$.
Homoplastically,
\begin{equation}\label{eq3.3.2}
  \delta' = M' \times \Delta,
  \parallel \delta'\parallel_{p} \leq \epsilon.
\end{equation}

The legal dataset $D_T$ and illegal dataset $D_F$ is generated by stamping the images in the clean dataset $\mathcal{D} $ with the different perturbation. Mathematically,
\begin{equation}\label{eq3.3.3}
  D_T = \{x + \delta \mid \forall  x \in \mathcal{D} \},
\end{equation}
\begin{equation}\label{eq3.3.4}
  D_F = \{x_i + \delta'_i \mid \forall  x_i \in \mathcal{D} \}.
\end{equation}
The perturbation is consistent $\delta$ in the legal dataset $D_T$ while the illegal dataset $D_F$ stamps differently random noise $\delta'_i$ for each image $x_i$.
In addition, there are two approaches to generate the label of the illegal dataset $D_F$ based on shuffling or not, namely, \emph{label-consistent} and \emph{label-inconsistent}.
The label-inconsistent approach is to shuffle the original label $L$ into $L'$.

\subsection{Model training}\label{3.4}

In the training phase, the number of images from $D_T$ and $D_F$ is equivalent per batch.
According to different approaches of dataset preprocessing, training methods can also divide into two categories, \emph{label-consistent}, and \emph{label-inconsistent}.
Equation \eqref{eq 3.3.1} is the loss function of label-consistent approach.
\begin{equation}\label{eq 3.3.1}
  \mathcal{L} = \mathcal{L}_1 + \lambda \mathcal{L}_2,
\end{equation}
where $\mathcal{L}_1=dist(\mathcal{F}(D_T),L)$, $\mathcal{L}_2 = - dist(\mathcal{F}(D_F),L)$. $dist(\cdot)$ is a metric, which in our scheme is cross entropy. Parameter $\lambda$ is used to control automatically the balance between two items.
Notice that $\mathcal{L}_2$ has an extra minus sign than $\mathcal{L}_1$.
Process of optimization aims to classify $D_T$ correctly while classifying $D_F$ incorrectly.

Likely, the loss function of the label-consistent approach is shown as \eqref{eq 3.3.2}.
\begin{equation}\label{eq 3.3.2}
  \mathcal{L} = \mathcal{L}_1 + \lambda \mathcal{L}_3,
\end{equation}
where $\mathcal{L}_3=dist(\mathcal{F}(D_F),L')$.
In train phase, the optimizer endeavors to classify $D_T$ correctly while classifying $D_F$ randomly.

\section{Performance analysis and results}\label{4}

We consider the classification model ResNet18 over the CIFER10 dataset.
The experimental equipment used is GeForce RTX 3090.
The the optimal accuracy  of ResNet18 is $P_{opt} = 92.68\%$, baseline $\mathcal{T} = 82.68\%$.

Low-complexity and flexibility are trivial for the proposed scheme.
Since the buyer only requires to stamp a specific perturbation before the image is fed into the ISNN, the computational complexity $\mathcal{O}_{buyer} = \mathcal{O} (1) \ll  \mathcal{O} (n)$.
Lightweight computation is sufficient and decryption is needless to access the model.
We limit $\epsilon \in [0,16]$ to ensure crypticity.
Other properties of the scheme are validated by experiments.

\subsection{Effectiveness and Fidelity}\label{4.1}

\begin{table}[htbp]
  \centering
  \begin{tabular}{@{}ccccccc@{}}
    \toprule
    \multirow{2}{*}{$\epsilon$} & \multicolumn{2}{c}{$D_T$} & \multicolumn{2}{c}{$D_F$} & \multicolumn{2}{c}{$\mathcal{D} $}                                                             \\ \cmidrule(l){2-7}
                         & \multicolumn{1}{c|}{lc}   & \multicolumn{1}{c|}{li}     & \multicolumn{1}{c|}{lc}   & \multicolumn{1}{c|}{li} & \multicolumn{1}{c|}{lc} & li    \\ \midrule
    1               & 90.68                     & 93.61                       & 2.01                      & 5.66                    & 2.38                    & 5.87  \\
    2               & 91.08                     & 92.51                       & 3.12                      & 11.95                   & 5.10                    & 11.62 \\
    4               & 92.29                     & 92.67                       & 4.96                      & 9.67                    & 4.21                    & 9.54  \\
    8               & 91.82                     & 92.70                       & 4.91                      & 8.35                    & 5.16                    & 8.38  \\
    16              & 90.11                     & 92.01                       & 0.44                      & 12.37                   & 0.56                    & 12.00 \\ \bottomrule
  \end{tabular}
  \caption{The performance of ISNN trained with different methods and perturbation intensity. lc: label-consistent method. li: label-inconsistent method.
  }  \label{tab4.1}
\end{table}

The accuracy of ISNN trained with two approaches is displayed in Table \ref{tab4.1}.
With enhancing perturbation intensity, the accuracy of the legal inputs of the ISNN trained with label-consistent method increases first and then decreases, reaching the peak value at perturbation intensity $ \epsilon=4$.
Meanwhile, the accuracy of models trained with the label-inconsistent method keeps decreasing.
In general, the performance of most classifiers $P_{buyer} > P_{opt} - 1\% > \mathcal{T}$, whereas all classifiers are keeping $P_{adv} < 15\% < \mathcal{T} $.
The performance of the classifier trained with label-consistent is weaker than label-inconsistent if the inputs are legal and stronger if inputs illegal, regardless of perturbation intensity is.

\subsection{Security and Verifiability}\label{4.2}
\begin{figure}[htbp]
  \begin{minipage}{0.48\linewidth}
    \centerline{\includegraphics[width=1\textwidth,trim= 135 20 210 20,clip]{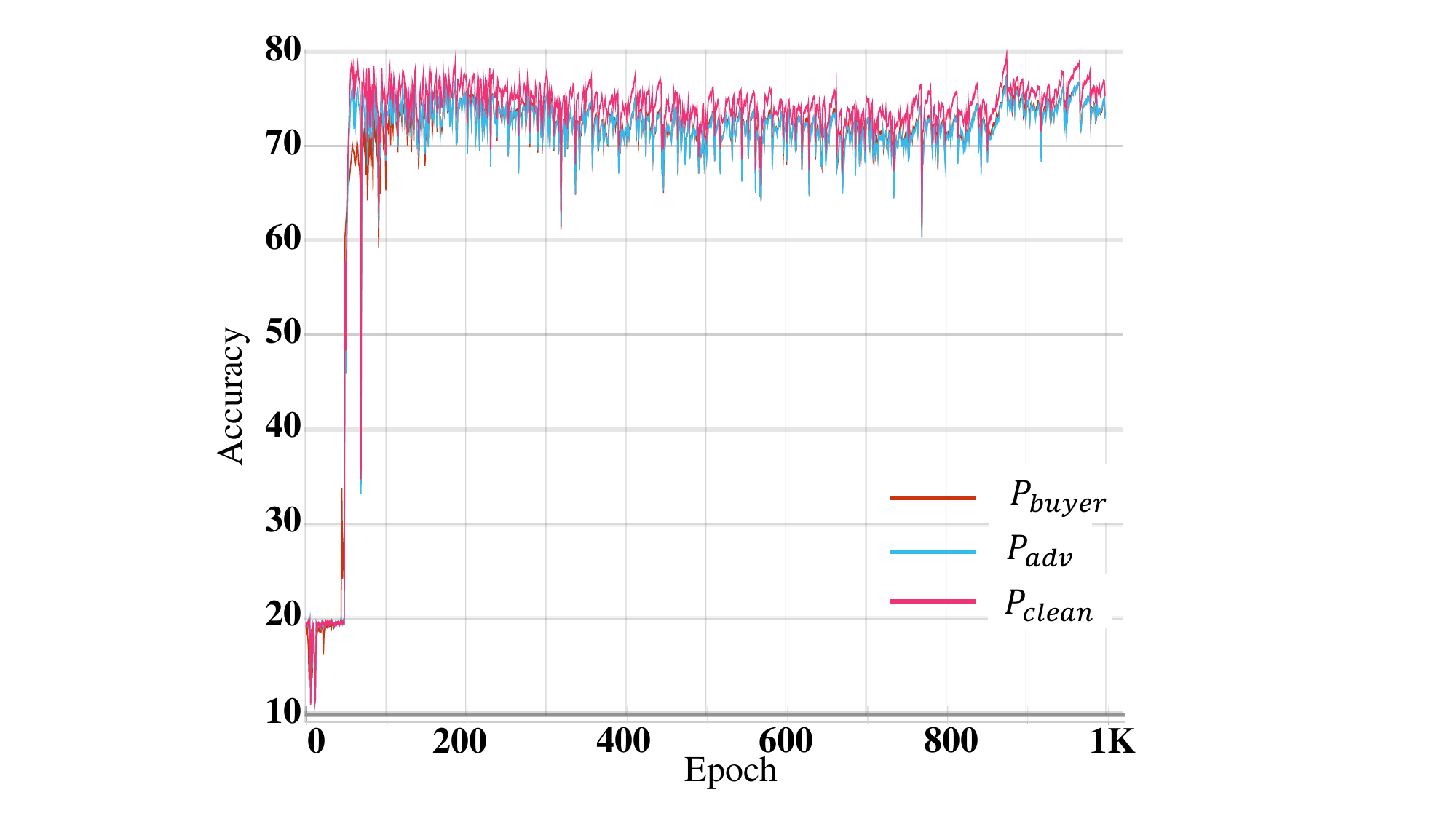}}
    \centerline{(a) dataset: $\mathcal{D}$, $\epsilon = 8$.}
  \end{minipage}
  \hfill
  \begin{minipage}{.48\linewidth}
    \centerline{\includegraphics[width=1.1\textwidth,trim= 165 35 190 40,clip]{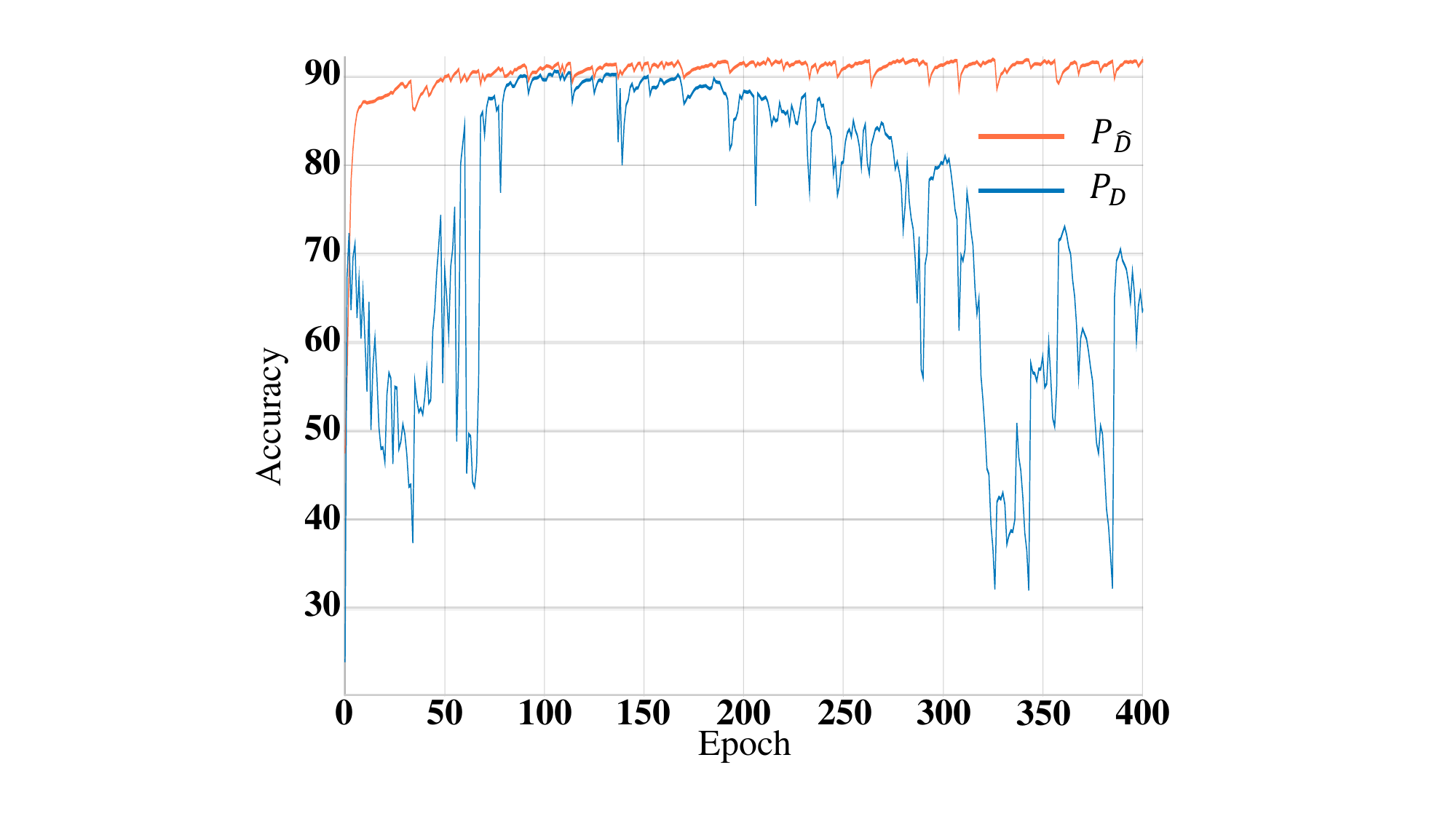}}
    \centerline{(b) dataset: $D_F$, $\epsilon = 4$.}
  \end{minipage}
\caption{The accuracy of the ISNN against retraining attack with different dataset, $\epsilon = 8$.}\label{fig4.1}
\end{figure}
\noindent \textbf{Retraining attack:}
Fig \ref{fig4.1}(a) shows the variation over the accuracy of the ISNN against retraining attack, where ISNN is ResNet18 trained with the label-consistent approach over 500 epochs.
Even after lengthy retraining with 1,000 epochs, the accuracy  $P_{adv}^{att} < 80\% < \mathcal{T} $.

\noindent \textbf{Forging copyright attack:}
Fig\ref{fig4.1}(b) illustrates the performance change process of ISNN for $D_T$ and $\widehat{D} $ as the training epochs increase.
Wherein, ISNN has trained 400 epochs with the label-consistent method.
The accuracy of $D_T$ remained high even after the same training epochs.
Therefore the retrain attack can successfully inject new copyright information into ISNN, but cannot erase the original copyright information.
\vspace{-0.2cm}
\begin{remark}
  The ISNN trained with label-inconsistent approach is not always effective against retraining attacks and forging copyright attack.
\end{remark}

\noindent \textbf{Reverse Iterating Attack:} 
Table \ref{tab4.2} shows the performance of models against iterating attacks.
The training method influences the attack success rate greatly, whereas perturbation intensity impacts the results weakly.
All ISNN trained with label-consistent has an attack success rate $P_{adv}^{att} < 75\% < \mathcal{T} $ regardless of the perturbation intensity is.
While the attack success rate $P_{adv}^{att} < 20\% < \mathcal{T} $ universally for label-inconsistent.
\begin{table}[htbp]
  \center
  \begin{tabular}{@{}cccccc@{}}
  \toprule
  $\epsilon$ & 1& 2& 4& 8 & 16\\ \midrule
  lc  & 72.35 & 74.91 & 73.15 & 74.16 & 72.41 \\
  li  & 16.88 & 16.91 & 18.01 & 16.88 & 17.24 \\ \bottomrule
  \end{tabular}\caption{Attack success rate of different training methods with various perturbation intensity.}\label{tab4.2}
\end{table}
\vspace{-0.35cm}
\section{Conclusion}\label{5}

This paper proposes a model copyright protecting scheme in the buyer-seller scenario based on ISNN, which does not require expensive computation overhead, nor does it need to encrypt and decrypt the model in the inference process.
The ISNN can be accessed by adding a specific perturbation to the input images.
However, once the perturbation added is forged, ISNN will return the false results.
Even if the adversary is allowed to access the entire dataset and  details except for the key, the adversary will not obtain satisfactory performance.
In conclusion, the proposed system is secure and effective.

\vfill\pagebreak
\bibliographystyle{IEEEbib}
\bibliography{refs}

\begin{thebibliography}{10}

\bibitem{onishi2019end}
Tadashi Onishi, Toshiyuki Motoyoshi, Yuki Suga, Hiroki Mori, and Tsuya Ogata,
\newblock ``End-to-end learning method for self-driving cars with trajectory
  recovery using a path-following function,''
\newblock in {\em 2019 International Joint Conference on Neural Networks
  (IJCNN)}. IEEE, 2019, pp. 1--8.

\bibitem{parkhi2015deep}
Omkar~M. Parkhi, Andrea Vedaldi, and Andrew Zisserman,
\newblock ``Deep face recognition,''
\newblock in {\em {BMVC}}, 2015, pp. 41.1--41.12.

\bibitem{szegedy2015going}
Christian Szegedy, Wei Liu, Yangqing Jia, Pierre Sermanet, Scott Reed, Dragomir
  Anguelov, Dumitru Erhan, Vincent Vanhoucke, and Andrew Rabinovich,
\newblock ``Going deeper with convolutions,''
\newblock in {\em Proceedings of the IEEE conference on computer vision and
  pattern recognition}, 2015, pp. 1--9.

\bibitem{RSArivest1978method}
Ronald~L Rivest, Adi Shamir, and Leonard Adleman,
\newblock ``A method for obtaining digital signatures and public-key
  cryptosystems,''
\newblock {\em Communications of the ACM}, vol. 21, no. 2, pp. 120--126, 1978.

\bibitem{AESdaemen1999aes}
Joan Daemen and Vincent Rijmen,
\newblock ``The rijndael block cipher: Aes proposal,''
\newblock in {\em First candidate conference (AeS1)}, 1999, pp. 343--348.

\bibitem{TDEScoppersmith1996proposed}
Don Coppersmith, Donald~Byron Johnson, and Stephen~M Matyas,
\newblock ``A proposed mode for triple-des encryption,''
\newblock {\em IBM Journal of Research and Development}, vol. 40, no. 2, pp.
  253--262, 1996.

\bibitem{Twofishknudsen1998deal}
Bruce Schneier, John Kelsey, Doug Whiting, David Wagner, Chris Hall, and Niels
  Ferguson,
\newblock ``Twofish: a 128-bit block cipher,''
\newblock {\em AES submission}, 1998.

\bibitem{chakraborty2020hardware}
Abhishek Chakraborty, Ankit Mondai, and Ankur Srivastava,
\newblock ``Hardware-assisted intellectual property protection of deep learning
  models,''
\newblock in {\em 2020 57th ACM/IEEE Design Automation Conference (DAC)}. IEEE,
  2020, pp. 1--6.

\bibitem{tian2021probabilistic}
Jinyu Tian, Jiantao Zhou, and Jia Duan,
\newblock ``Probabilistic selective encryption of convolutional neural networks
  for hierarchical services,''
\newblock in {\em Proceedings of the IEEE/CVF Conference on Computer Vision and
  Pattern Recognition (CVPR)}, 2021, pp. 2205--2214.

\bibitem{guo2018watermarking}
Jia Guo and Miodrag Potkonjak,
\newblock ``Watermarking deep neural networks for embedded systems,''
\newblock in {\em 2018 IEEE/ACM International Conference on Computer-Aided
  Design (ICCAD)}. IEEE, 2018, pp. 1--8.

\bibitem{hitaj2018have}
Dorjan Hitaj and Luigi~V. Mancini,
\newblock ``Have you stolen my model? evasion attacks against deep neural
  network watermarking techniques,''
\newblock {\em CoRR}, vol. abs/1809.00615, 2018.

\bibitem{le2020adversarial}
Erwan Le~Merrer, Patrick Perez, and Gilles Tr{\'e}dan,
\newblock ``Adversarial frontier stitching for remote neural network
  watermarking,''
\newblock {\em Neural Computing and Applications}, vol. 32, no. 13, pp.
  9233--9244, 2020.

\bibitem{darvish2019deepsigns}
Bita Darvish~Rouhani, Huili Chen, and Farinaz Koushanfar,
\newblock ``Deepsigns: An end-to-end watermarking framework for ownership
  protection of deep neural networks,''
\newblock in {\em Proceedings of the Twenty-Fourth International Conference on
  Architectural Support for Programming Languages and Operating Systems}, 2019,
  pp. 485--497.

\bibitem{uap}
Seyed-Mohsen Moosavi-Dezfooli, Alhussein Fawzi, Omar Fawzi, and Pascal
  Frossard,
\newblock ``Universal adversarial perturbations,''
\newblock in {\em Proceedings of the IEEE conference on computer vision and
  pattern recognition (CVPR)}, 2017, pp. 1765--1773.

\bibitem{MPC}
Andrew Chi-Chih Yao,
\newblock ``How to generate and exchange secrets,''
\newblock in {\em 27th Annual Symposium on Foundations of Computer Science
  (SFCS)}. IEEE, 1986, pp. 162--167.

\bibitem{HE}
Craig Gentry,
\newblock ``Fully homomorphic encryption using ideal lattices,''
\newblock in {\em Proceedings of the forty-first annual ACM symposium on Theory
  of computing}, 2009, pp. 169--178.

\bibitem{SS}
Adi Shamir,
\newblock ``How to share a secret,''
\newblock {\em Communications of the ACM}, vol. 22, no. 11, pp. 612--613, 1979.

\end{thebibliography}

\end{document}